\documentclass{article}
\usepackage[utf8]{inputenc}
\usepackage[ruled,vlined]{algorithm2e}
\usepackage{graphicx}
\usepackage{float}
\usepackage{amssymb,latexsym,amsfonts}
\usepackage{amsmath,amssymb}
\usepackage{fancyhdr}
\usepackage{color}
\usepackage{colortbl}
\title{A new algorithm for graph center computation and graph partitioning according to the distance to the center}
\author{Fr\'ed\'eric Protin}
\date{2 October 2019}

\begin{document}

\maketitle

\noindent Abstract: {\it{
We propose a new algorithm for finding the center of a graph, as well as the rank of each node in the hierarchy of distances to the center. In other words, our algorithm allows to partition the graph according to nodes distance to the center. Moreover, the algorithm is parallelizable. We compare the performances of our algorithm with the ones of Floyd-Warshall algorithm, which is traditionally used for these purposes. We show that, for a large variety of graphs, our algorithm outperforms the Floyd-Warshall algorithm.}}

\section{Introduction}

The \textit{center} of a non-oriented graph, also called its Jordan center, is the set of all vertices of minimum eccentricity, that is, the set of all vertices for which the greatest distance to other vertices is minimal. Equivalently, it is the set of vertices with eccentricity equal to the graph's radius. Vertices belonging to the center are called central nodes. (See e.g. \cite{G} for precise definitions).
\\

Finding the center of a graph is useful e.g. in facility location problems, where the goal is to minimize the worst-case distance to the facility. For example, placing a fire station at a central point reduces the longest distance the fire truck has to travel. \cite{Luo} showed that the center of a graph related to a rumor spread is a good estimator for the rumor source (see also \cite{K}). We are also interested in partitioning the graph according to nodes distance to the center. This partitioning is useful, for example, for visualization or reporting, as showed in Figure 1 below, where the graph represents the cable network between electronic devices, as proposed in the Connect'it project of the enterprise Activus-group. Vincent Viton of Activus-group (\cite{PC}) suggested to use the nodes hierarchy in order to improve visualization.\\

\begin{figure}
\begin{center}

\includegraphics[scale=0.23]{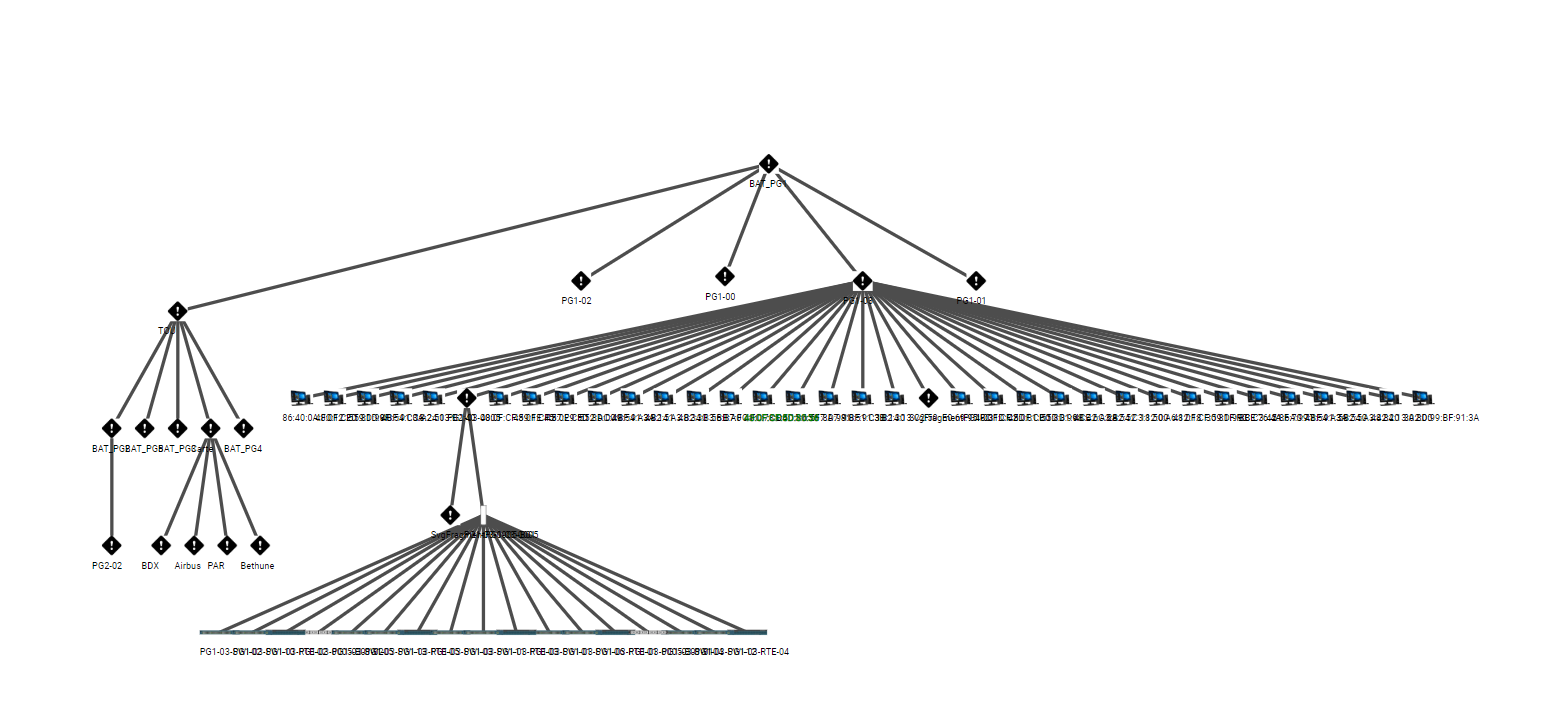}
\caption{Application of nodes hierarchization of a graph to visualization. In this example the graph is a tree.}
\end{center}
\end{figure}

For a non-oriented graph $G=(V,E)$, where $V$ is the set of nodes and $E$ the set of edges, Bellman-Ford algorithm \cite{B,Fo,M} computes the distance from a fixed node to every other node. Its complexity is ${\displaystyle O(|V||E|)}$, where $|\cdot|$ denotes the cardinal of a set. Therefore, the complexity in the worse case, i.e. for a dense graph, is ${\displaystyle O(|V|^3)}$. Dijkstra's algorithm \cite{D}, applied to the same graph, returns in addition the paths from the fixed node to every other node. When applied to our context of interest, these algorithms return more than the required information, at the price of being slower than the algorithm we propose in the present work. Note that another algorithm was proposed in \cite{H} for trees.

\section{Principles of the new algorithm}

Let $G = (V,E)$ be a connected graph, with $V$ being the set of nodes (or vertices), and $E$ the set of edges. Recall that the {\it{center}} of the graph $G$ is the set of nodes that minimizes the greatest distance to the other nodes. Here we consider the distance $\text{dist}(\cdot, \cdot):V\times V\to \mathbb{N}$, where the distance $\text(i,j)$ between two nodes $i$ and $j$ is the smallest number of edges in a path joining them. Therefore, the center of $G$ is the set $$\displaystyle \mathcal{C}_0:=\Big\{i\in V :  \max_{j}{\text{dist}(i,j) }= \min_{i}{\max_{j}{\text{dist}(i,j)\Big\}}}.$$

Besides determining the center of a graph, we also want to obtain a partition of the graph according to nodes distance to the center : $$\displaystyle G = \bigsqcup_{n\in\mathbb{N}} \Big\{i\in V : \max_j \text{dist}(i,j) =\min_i\max_j \text{dist}(i,j)+n\Big\}=: \bigsqcup_{n\in\mathbb{N}}\mathcal{C}_n.$$

Let $A$ denote the adjacency matrix of the graph $G$. We denote by $\tilde{A}$ the matrix obtained from $A$ by changing the diagonal coefficients into 1, i.e. $\tilde{A}_{i j}=A_{i j}$ for $i\neq j$, and $\tilde{A}_{i i}=1$ for every $i$. In other words, $\tilde{A}$ is the adjacency matrix of the graph $\tilde G$, obtained from $G$ by imposing the fact that every node has a single edge connecting it to itself.

Note that, for every $i,j \in V$, the coefficient $(\tilde{A}^n)_{i j}$ of the $n$th power of the matrix $\tilde{A}$ indicates the number of paths of length at most $n$ going from node $i$ to node $j$. 

Note also the fact that, if for given $i,j \in V$, $n$ is the smallest integer such that $(\tilde{A}^n)_{i j}\neq 0$, then the shortest path from $i$ to $j$ has length equal to $n$. 

We deduce the following method for computing the center of the graph and for partitioning it.\\

\paragraph{The main idea of the algorithm}$ $\\

\noindent Compute the successive powers $\tilde{A}^n$ of the matrix $\tilde{A}$, for $n \geq 1$, until at least one row gets filled with non-null coefficients. More precisely, let us denote by $n_0$ the smallest power such that the matrix $\tilde{A}^{n_0}$ contains at least one row with no null coefficients:
$$
n_0 := \min\left\{n \geq 1 : \exists \ i \in V \text{ s.t. } (\tilde{A}^{n})_{i j} \neq 0, \ \forall j \in V\right\}.
$$
Note that the indexes of those rows in $\tilde{A}^{n_0}$ correspond to the center nodes of the graph and hence form the set $\mathcal{C}_0$:
$$
\mathcal{C}_0 := \left\{i \in V  : (\tilde{A}^{n_0})_{i j} \neq 0, \ \forall j \in V\right\}. 
$$

In order to completely partition the graph and determine $\mathcal{C}_k$, for every $k \geq 1$, we continue computing successively the powers $\tilde{A}^n$, for $n = n_0 + k, \ k \geq 1$, until all the rows of $\tilde{A}^n$ get filled with non-null coefficients. 

Note that for any $k \geq 1$, the set $\mathcal{C}_k$ is formed by the indexes of the rows having at least one null coefficient in $\tilde{A}^{n_0+k-1}$, but no null coefficients in  $\tilde{A}^{n_0+k}$:
$$
\mathcal{C}_k := \left\{i \in V \setminus \mathcal{C}_{k-1} : (\tilde{A}^{n_0+k})_{i j} \neq 0, \ \forall j \in V\right\}. 
$$
Note that, since the graph is connected, we have $\mathcal{C}_k \neq \emptyset$ for $k=0,\ldots,r$, and $\mathcal{C}_k = \emptyset$ for $k > r$, where $r$ is the radius of the graph.

\paragraph{An improved version}$ $\\

\noindent The method presented above solves the problem of finding the center of a graph and of partitioning the graph, but it is expensive in time. We present in the following few improvements which significantly diminish the computing time, resulting in a quite powerful algorithm. \\

\noindent \textbf{Improvement 1.} Note that the only important thing to know about the coefficients of $\tilde{A}^n$, when applying the previous method, is if they vanish or not. Therefore, the method still works if the non-null coefficients in the matrix $\tilde{A}$, as well as in its powers, are replaced by $1$. Let us formalize this remark, and show how we can use it in order to accelerate the previously described algorithm. 

Denote by $M\mapsto R(M)$ the application which replaces each of the coefficients $m_{ij}\in \mathbb{N}$ of a matrix $M$ by $\max{(m_{ij}, 1)}$. Note that this operation commutes with the matrix product. Consequently, if we want to compute $R(AB)$ for two matrices $A$ and $B$ with coefficients in $\mathbb{N}$, we can instead prefer to compute the product $R(A)R(B)$, which gives the same result and faster. Indeed, when we multiply a row $(\ell_1, ..., \ell_k)$ by a column $(c_1, ..., c_k)^T$, we can stop the recursive computation of the scalar product as soon as $\ell_i=c_i=1$ for some index $i$, and return the value $1$.\\

\noindent \textbf{Improvement 2.} Since the coefficient $(\tilde{A}^n)_{i j}$  of $\tilde{A}^n$ equals  the number of paths of length at most $n$ from node $i$ to node $j$, the sequence $((\tilde{A}^n)_{i j})_n$ is increasing. Thus $(\tilde{A}^{n-1})_{i j} \geq 1$ entails $(\tilde{A}^{n})_{i j}\geq 1$. It is therefore useless to compute a coefficient $(\tilde{A}^n)_{i j}$ when $(\tilde{A}^{n-1})_{i j}\geq 1$. \\

\noindent \textbf{Improvement 3.} A naive way to determine $n_0$ is to compute successively the terms of the sequence $(\tilde{A}^n)_n$, until one row gets filled with non-null coefficients. We propose here a faster way to determine $n_0$, as follows. 

We compute recursively the sequence of powers $\tilde{A}, \tilde{A}^2, \tilde{A}^4, \tilde{A}^{16}, \ldots,$ $ \tilde{A}^{2^n},\ldots$, until at least one row gets filled with non-null coefficients, say for some $\tilde{A}^{2^m}$, with $m \geq 1$. We form a list with all these powers of $\tilde{A}$ until $\tilde{A}^{2^{m-1}}$ .

We then construct recursively a subset $\mathcal{A}$ of $\{ 1,2,\ldots,m-1 \}$, in the following manner. We first initialize $\mathcal{A} := \{m-1\}$. We then consider one by one the numbers $m-2, m-3,\ldots,2,1$ in this order, and add to the current subset $\mathcal{A}$ those numbers $k$ which satisfy the fact that the product $\displaystyle \tilde{A}^{2^{k}}\cdot \prod_{\ell \in \mathcal{A}, \ \ell > k} \tilde{A}^{2^{\ell}}$ does not contain any row filled with non-null coefficients.

It is straightforward that $$\displaystyle\tilde{A}^{n_0-1}=\displaystyle\prod_{\ell \in \mathcal{A}}\tilde{A}^{2^{\ell}},$$
and $\displaystyle n_0-1=\sum_{\ell \in \mathcal{A}}2^{\ell}$, which is the unique binary decomposition of $n_0-1$.

$ $\\

\noindent \textbf{Improvement 4.} Since the graph is non oriented, the matrix $\tilde{A}^n$ is symmetric and thus the computing time of $\tilde{A}^n$ is reduced by half.\\

A detailed description of the final algorithm using the above improvements is given in the next section.

\section{Description of the algorithm}

In this section we present in detail the final algorithm, which takes into account the main ideas and the improvements previously presented. 

Improvements 1, 2 and 4 lead to the following function (called \textit{multiply}) for computing $R(M\cdot M2)$, the product of two matrices $M$ and $M2$, composed with the application $R$ described in Improvement 2, which replaces each coefficient by its maximum with $1$. 
This function will be used for computing the powers $\tilde{A}^n$.
The variable $width$ represents the number of rows of the matrix $M$.

\begin{algorithm}[H]
\SetAlgoLined

$M3 \leftarrow M2$\\
\For{$\ell \gets 0$ to $width$}{
\For{$c \gets l+1$ to $width$}{
\If{$M3[\ell][c] =0$}{
\For{$i \gets 0$ to width}{
\If{$M[\ell][i] \geq 1$ and $M2[i][c] \geq 1$}{
$M3[\ell][c] = 1$\\
$M3[c][\ell] = 1$\\
break
}
}
}
}
}
return $M3$
 \caption{def multiply(M : matrix, M2 : matrix, width : int):}
\end{algorithm}  

$ $\\

We will also use the following variant (called \textit{multiplyIns}) which performs the same multiplication as before, but also inserts the sets $\mathcal{C}_k, k \geq 0$ into a dictionary, as the algorithm progresses. The keys of this dictionary are the nodes of the graph, and the values are the corresponding distances to the center of the graph.

\begin{algorithm}[H]
\SetAlgoLined
$M3 \leftarrow M2$\\
\For{$\ell \gets 0$ to width}{
\For{$c \gets \ell+1$ to width}{
\If{$M3[\ell][c] = 0$}{
\For{i $\gets 0$ to width}{
\If{$M[\ell][i] \geq 1$ and $M2[i][c] \geq 1$}{
$M3[\ell][c] = 1$\\
$M3[c][\ell] = 1$\\
\tcc{$rows[\ell]$ is the number of non-null coefficients of the $\ell$th row of M3.}
$rows[\ell]\leftarrow rows[\ell]+1$\\
$rows[c]\leftarrow rows[c]+1$\\
\If{$rows[\ell] = width$}{
$dictNodes[nodes[\ell]] \leftarrow round$ \tcp*{insert data.}
}
\If{$lines[c] = width$}{
$dictNodes[nodes[c]] \leftarrow round$ \tcp*{insert data.}
}
}
}
}
}
}
return $M3$\
 \caption{def multiplyIns(M : matrix, M2 : matrix, width : int, round : int, rows : list of int):}
\end{algorithm}  

$ $

The last function we use (called \textit{testRow}) checks, for a given matrix $M$, if it contains any row filled with non-null coefficients.

$ $

\begin{algorithm}[H]
\SetAlgoLined
\KwResult{def testRow(M, width): }
$output \leftarrow FALSE$ \\
 \For{$\ell \gets 0$ to width}{
    $ s\leftarrow |\{x \in M2[\ell]:\text{ } x \neq 0\}|$\tcp*{Number of coefficients $\neq 0$.}
    \If{s $=$ width}{
    $output \leftarrow TRUE$\\
   break
    }

  }
return $output$
\caption{def testRow(M : matrix, width : int):}
\end{algorithm}  

$ $

The final algorithm (called \textit{Partitioning}) is then the following. Here MAdjacency denotes the adjacency matrix of the graph, and $width$ its number of rows, which are given as input. Identity matrix of order $n$ is denoted Identity($n$).

\begin{algorithm}[H]
\SetAlgoLined
\tcc{Initialization.}
$M \leftarrow MAdjacency + Identity(width)$ \tcp*{The matrixList $\tilde{A}$.}
$M2 \leftarrow M$\\
$matrixList \leftarrow [M]$ \tcp*{list initialized with the matrix $M$.}

\tcc{Computation of the list of successive powers $\tilde A^{2^n}$. }

\For{$round \gets 0$ to width}{
    $product \leftarrow multiply(matrix[-1], matrix[-1], width)$\\
    \If{ (product, width)\tcp*{Check if a row is filled.}}{
    break
    
    }
    
    $matrixList.add(product)$\
  }

\tcc{Computation of $\tilde A^{n_0-1}$.}
\If{length(matrixList) $\geq 2$}{
$encode \leftarrow length(matrixList)-1$\\
$aux \leftarrow matrixList[-1]$\\
\For{round $\gets 0$ to width}{
$U \leftarrow multiply(aux, matrixList[encode$-1$], width)$\\
\If {not testRow(U, width)\tcp*{Check if a row is filled.}}{
$aux \leftarrow U$\\
}
$encode \leftarrow encode -1$\\
\If{$encode = 0$}{
break
}          
}
$M2 \leftarrow aux$\\
}
\tcc{Counting the number of non-null coefficients in each row}

\For{$\ell \gets 0$ to width}{
$rows[\ell] \leftarrow|\left\{x \in M2[l]: x \neq 0\right\}|$\tcp*{Nb. of coefficients $\neq 0$.}

\tcc{If the matrix $\tilde A$ contains a row without null coefficients, fill the dictionary with center nodes.}
\If{$rows[\ell] =$ width}{
$dictNodes[nodes[l]] = 0$ 
}

}
\tcc{Computation of the complete hierarchy of nodes.}
\For{round $\gets 0$ to width}{

\If{ length(dictNodes) $=$ width}{

        break
        }


    $M2 = multiplyIns(M, M2, width, round, rows) $
    
}

print($dictNodes$) \tcp*{print the final result}

\caption{Partitioning(MAdjacency : matrix, width : int)}
\end{algorithm} 

\noindent {\bf{Scalability}}. Each step described in Section 1 is obviously parallelizable. The proposed algorithm is therefore parallelizable in a straightforward manner.

\section{Experimental results and conclusions}

An explicit derivation of the computational complexity of the algorithm is a quite tedious question, because it depends highly on the geometry of the graph.

We choose instead to compare experimentally the algorithm we propose with the Floyd-Warshall algorithm. Note that, by construction, the computing time of the algorithm we propose decreases when the connectivity of the graph increases, or equivalently, when the depth of the graph decreases. The worse case for the algorithm we propose is when the graph is linear. In this case, the Floyd-Warshall algorithm is faster than our new algorithm, but still the computing times are of the same order.

The experiments were performed using the library {\it{Networkx}} in Python. We generated random graphs with a given morphology (number of nodes (N), number of edges (NA) and depth (P)). Recall that the depth of a graph is the maximal distance from a node to the center. 

In Table \ref{table} we report an extract of our experiments. The computing times for the Floyd-Warshall algorithm and for the algorithm we propose are reported in seconds. Unfavorable cases for our proposed algorithm are highlighted in red, and improvements of a factor $\geq 10$ for the computing time with respect to the Floyd-Warshall algorithm are highlighted in green. \\

\noindent {\bf{Conclusion.}} The experiments performed suggest that the algorithm we propose is faster that the Floyd-Warshall algorithm, except for graphs with a high ratio $P/N$.\\

Note also that the Improvement 3 is optional, but beneficial. We have experimentally noted that, as we could expect, this improvement slows down the algorithm a little bit on shallow graphs, but accelerates it a lot on very deep graphs.\\

\begin{table}
\caption{Comparison of computing time (in seconds) of the new algorithm and the Floyd-Warshall algorithm, for different graph morphologies.}
\begin{center}

\begin{tabular}{|c|c|c|}\label{table}
\begin{bf}Graph morphology\end{bf} & \begin{bf}Floyd-Warshall\end{bf} & \begin{bf}New algorithm\end{bf} \\
\hline
N= 500 , NA= 995, P = 15 & 189 & 114\\
\hline
N = 500 , NA = 1495, P = 17 & 198 & 113\\
\hline
N = 500 , NA = 1491, P = 14 & 191 & 97\\
\hline
N = 500 , NA = 1981, P = 14 & 195 & 82\\
\hline
N = 500 , NA = 2477, P = 14 & 201 & 89\\
\hline
N = 500 , NA = 2962 , P = 14 & 194 & 76\\
\hline
N= 500 , NA= 5365 , P = 13 & 186 & 62\\
\hline
N= 500 , NA= 14511, P = 13 & 196 & 55\\
\hline
N= 500 , NA= 22840 , P = 13 & 186 & 49\\
\hline
N= 500 , NA= 40824, P = 13 & 198 & 52\\
\hline
\rowcolor{green}N= 500 , NA= 14632, P = 3 & 203 & 13\\
\hline
N= 500 , NA= 14515, P = 8 & 191 & 26\\
\hline
\rowcolor{green}N= 500 , NA= 14569,  P = 6 & 193 & 18\\
\hline
\rowcolor{green}N= 500 , NA= 14556, P = 3 & 193 & 10\\
\hline
\rowcolor{green}N= 500 , NA= 14550, P = 2 & 204 & 8\\
\hline
\rowcolor{green}N= 500 , NA= 14553, P = 1 & 189 & 10\\
\hline
N= 500 , NA= 14402, P = 25 & 198 & 147\\
\hline
N= 500 , NA= 22563, P = 25 & 192 & 132\\
\hline
N= 500 , NA= 2964, P = 26 & 192 & 157\\
\hline
N= 500 , NA= 2963, P = 4 & 196 & 54\\
\hline
N= 200 , NA= 1161, P = 3 & 12 & 2\\
\hline
N= 200 , NA= 2915, P = 8 & 12 & 3\\
\hline
\rowcolor{red}N= 200 , NA= 1473, P = 75 & 13 & 51\\
\hline
\rowcolor{red}N= 400 , NA= 15075, P = 75 & 98 & 383\\
\hline
N= 400 , NA= 17412, P = 25 & 97 & 90\\
\hline
N= 400 , NA= 4247, P = 26 & 99 & 85\\
\hline
N= 400 , NA= 4276, P = 16 & 97 & 43\\
\hline
N= 400 , NA= 30467, P = 15 & 97 & 48\\
\hline
\rowcolor{green}N= 400 , NA= 31636, P = 2 & 100 & 1\\
\hline
\rowcolor{green}N= 400 , NA= 31329, P = 6 & 100 & 10\\
\hline
N= 400 , NA= 17807, P = 7 & 98 & 12\\
\hline
N= 400 , NA= 7963, P = 7 & 100 & 13\\
\hline
\rowcolor{green}N= 400 , NA= 7964, P = 3 & 102 & 6\\
\hline
N= 400 , NA= 4270, P = 5 & 100 & 15\\
\hline
N= 600 , NA= 6460, P=6 & 338 & 60\\
\hline
N= 800 , NA= 8674, P =6 & 794 & 141\\
\hline
\rowcolor{green}N= 800 , NA= 311332, P=5 & 793 & 37\\
\hline
\rowcolor{green}N= 1000 , NA= 39301, P = 4 & 1503 & 81\\
\hline
N= 1000 , NA= 39164, P = 25 & 1403 & 558\\
\hline
N= 1000 , NA= 10868, P = 26 & 1510 & 727\\
\hline
\rowcolor{red}N= 200 , NA= 1998, P=26 & 11 & 17\\

\hline
\rowcolor{red}N= 300 , NA= 3145, P = 26 & 41 & 42

\end{tabular}

\end{center}
\end{table}


\end{document}